# Anomalous tensile strength and thermal expansion, and low thermal conductivity in wide band gap boron monoxide monolayer

Bohayra Mortazavi*,[a], Fazel Shojaei**,[b], Fei Ding[c] and Xiaoying Zhuang***,[a,d]

[a]Chair of Computational Science and Simulation Technology, Department of Mathematics and Physics, Leibniz Universität Hannover, Appelstraße 11,30167 Hannover, Germany.
[b]Department of Chemistry, Faculty of Nano and Bioscience and Technology, Persian Gulf University, Bushehr 75169, Iran.
[c]Institut für Festkörperphysik, Leibniz Universität Hannover, Appelstraße 2, 30167, Hannover, Germany.
[d]College of Civil Engineering, Department of Geotechnical Engineering, Tongji University, 1239 Siping Road Shanghai, China.

**Abstract**

Most recently the formation of boron monoxide (BO) in the two-dimensional (2D) form has been confirmed experimentally (*J. Am. Chem. Soc.* 2023, 145, 14660). Motivated by the aforementioned finding, herein we theoretically explore the key physical properties of the single-layer and suspended BO. Density functional theory (DFT) results reveal that BO monolayer yields a large indirect band gap of 3.78 (2.18) eV on the basis of HSE06(PBE) functional. Ab-initio molecular dynamics results reveal the remarkable thermal stability of the BO monolayer at 1000 K. The thermal and mechanical properties at room temperature are furthermore investigated using a machine learning interatomic potential (MLIP). The developed MLIP-based model close to the ground state could very precisely reproduce the DFT predictions for the mechanical properties of the BO monolayer. The elastic modulus, tensile strength and lattice thermal conductivity of the BO monolayer at room temperature are predicted to be 107 GPa, 25 GPa and 5.6±0.5 W/mK, respectively. At the room temperature the BO monolayer is noticeably predicted to yield an ultrahigh negative thermal expansion coefficient, by almost 17 folds larger than that of the single-layer graphene. The presented results reveal the large indirect electronic band gap, decent thermal and dynamical stability, anomalously low elastic modulus to tensile strength ratio, ultrahigh negative thermal expansion coefficients and low lattice thermal conductivity of the BO monolayer.

1. Introduction

Wartik and Apple [1] almost seven decades ago reported the first synthesis of a novel material with an empirical formula of boron monoxide (BO), using the condensation of tetrahydroxydiboron molecule at 400 °C. Because of the hard and brittle nature of the fabricated BO lattices, its structural features have not been precisely determined until very recently. Motivated by the latest emergences of the boron-based nanomembranes such as the insulating hexagonal boron nitride and metallic borophene lattices [2–4], Perras and coworkers [5] most recently conducted elaborated NMR measurements to shed light on the atomic structure of the BO system, synthesized by the thermal treatment of tetrahydroxydiboron in a tube furnace. Their detailed analysis confirms that the experimentally realized BO samples are consisting of the 2D lattices with a random stacking pattern. On this basis, likely to the case of the graphene [6,7], the mechanical exfoliation technique can be also employed to realize the single-layer and suspended form of the bulk BO system.

From the practical point of view, it is very important to evaluate the stability and key physical properties of the BO nanosheets to elaborate their suitability for various applications. In order to address the aforementioned necessity, for the first time herein a combination of first-principles density functional theory (DFT) calculations and machine-learning-based classical modeling were conducted to evaluate the stability and explore the electronic, mechanical, thermal expansion and lattice thermal conduction properties of the suspended and defect-free BO monolayer. First-principles results confirm the outstanding thermal and dynamical stability of the suspended BO monolayer. Analysis of the electronic band structure with the HSE06 hybrid functional reveals that the BO monolayer is a semiconductor with an indirect band gap of 3.78 eV, which yields decent optical absorption only near the UV region of light. The room temperature thermal conductivity of the BO monolayer is estimated to be 5.6 ±0.5 W/mK, which is noticeably low. As an interesting finding, the thermal expansion coefficient of the suspended BO monolayer at room temperature is found to be by almost 17 folds larger than that of the graphene. Moreover, the elastic modulus to ultimate tensile strength ratio of the BO monolayer at room temperature is predicted to be around 4, which is almost half of that for graphene and other prominent 2D materials. The results provided by the state-of-the-art theoretical calculations, reveal the interesting/anomalous electronic, mechanical and thermal properties of the boron monoxide nanosheets, highly appealing for more elaborated experimental and theoretical endeavours.

## 2. Computational methods

Vienna ab initio simulation package (VASP) [8,9] was employed for DFT calculations with Perdew-Burke-Ernzerhof (PBE) within the generalized gradient approximation (GGA), Grimme's DFT-D3 [10] van der Waals (vdW) dispersion correction and a kinetic energy cutoff of 500 eV. Optimization of both lattice parameters and atomic positions were conducted using the conjugate gradient method adopting a 11×11×1 Monkhorst-Pack [11] k-point grid and energy and force convergence criteria of $10^{-5}$ eV and 0.002 eV/Å, respectively. To prevent vdW interactions along the thickness, a 15 Å vacuum spacing was employed. Ab-initio molecular dynamics (AIMD) simulations were carried out at 500 and 1000 K with a time step of 1 fs for 10 ps, to inspect the thermal stability. A moment tensor potential (MTP) [12] was fitted to investigate the phononic, thermal and mechanical properties with the aid of MLIP package [13]. The training dataset was prepared by the AIMD calculations over unitcells with a time step of 1 fs, DFT-D3, and 2×2×1 grid over stress-free and uniaxially loaded systems under varying temperatures, similar to our recent works [14,15]. A MTP with a cutoff distance of 3.5 Å was trained using the two-step passive training approach [16] as that of our recent studies [14,15]. The phonon dispersion relation was obtained using the developed MTP, over a 5×5×1 supercell and by adopting the small displacement method of the PHONOPY package [17], as explained in our earlier investigation [18]. We utilized the LAMMPS package [19] to examine thermal and mechanical properties based on the developed MTP. Mechanical response was examined using the quasi-static uniaxial tensile loading [14,15]. Non-equilibrium molecular dynamics (NEMD) simulations were carried to evaluate lattice thermal conductivity of the BO monolayer, using the same approach as that developed in our recent studies [20–22].

## 3. Results and discussions

We first investigate the structural features of the BO monolayer, as it is illustrated in Fig. 1a. The BO monolayer is a topographically completely flat structure, which shows a square lattice with the space group of P4/mmm (No. 123) and an optimized lattice constant of 7.83 Å. This nanosheet can be thought of $B_8O_4$ moieties, which structurally resembles the crown ether of 12-crown-4, covalently bounded through oxygen bridges. The size of pores in BO monolayer, however, is greater than that of 12-crown-4. The BO sheet is nanoporous and like the crown ethers might be a suitable candidate for the ion sensing [23]. The layered bulk BO can be synthesized by thermal condensation of tetrahydroxydiboron molecules [1,5]. To examine the thermodynamic favorability of the formation of BO monolayer from the tetrahydroxydiboron molecules, we calculated the reaction energy of tetrahydroxydiboron condensation according to the reaction $4B_2O_4H_4 \rightarrow B_8O_8$ (primitive cell) $+ 4H_2O$. It is found that the reaction is by +2.35 eV/unitcell endothermic and thus thermodynamically unfavorable, which may explain the need of 400 °C high temperature for the synthesis of the layered bulk BO. The bond distance obtained for boron pairs is 1.74 Å, which is within the range of 1.61-1.88 Å, previously obtained for sB- and β12-borophenes [23,24]. The B-O bond length, however, is found to be 1.37-1.39 Å, slightly larger than that in boronic acid (1.36 Å). To examine the nature of interactions in BO monolayer, electron localization function (ELF) [25] is calculated and shown in Fig. 1b. The presence of large electron localization around the middle of bonds indicates that the B-B and B-O bonds in this system are covalent.

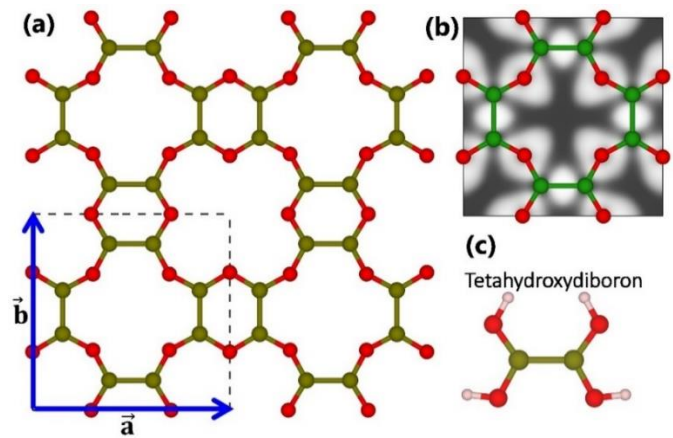

Fig. 1. (a) Top view of the BO monolayer's flat crystal structure. (b) The ELF contour map on the sheet's plane ranging from 0 (dark) to 1 (fully-bright) and (c) depicts the molecular geometry of tetahydroxydiboron molecule. Green, red, and pink balls in this figure indicate boron, oxygen, and hydrogen atoms. These results are plotted using the VESTA [26] software.

In order to investigate the electronic properties of the BO monolayer, we carried out the electronic band structure calculations using PBE and HSE06 functionals. Fig. 2 shows the HSE06 band structure, orbital projected band structure, atom-type partial density of states (PDOS), and charge density distribution of valence band maximum (VBM) and conduction band minimum (CBM) of the suspended single-layer BO. The BO monolayer is found to be a wide gap semiconductor with an indirect band gap of 3.78 (2.18) eV according to the HSE06(PBE) functional, in which the VBM and CBM are located ate Γ and S points, respectively. This band gap is appreciably smaller than the quasi-direct band gap of 6.50 eV obtained for the other boron-contained binary 2D material, h-BN [27], while it is larger than the direct gap of 1.51 eV obtained for the widely studied black phosphorene [28]. Both valance and conduction bands show narrow dispersions. The orbital projected band structure indicates that the valance states within the energy range show in Fig. 2 are in-plane states made of $p_x$ and $p_y$ orbitals, while the conduction states are p-states made of $p_z$ orbitals. PDOS and charge density distributions of band edge states show that VBM is contributed by both B and O atoms, representing primarily σ(B-B) state hybridized with in-plane side overlap between B and O atoms. The CBM, however, is only localized on isolated B pairs, representing π(B-B) states. The almost localized character of both VBM and CBM may explain the narrow dispersions of valance and conduction bands. Likely to the crown ethers with similar nanoporous structures, the wide band gap BO monolayer might also be a suitable candidate for ion sensing and separation [23,29,30]. We also investigated the optical properties of BO monolayer by calculating the frequency dependent absorption coefficients using the HSE06 functional, as shown in Fig. S1 of the supporting information document. Due to the symmetry, the absorption spectrum of BO is independent of the light polarization along the in-plane directions of [100] and [010]. According to Fig. S1, the first absorption peak of the BO monolayer appears at 6.67 eV which is in the ultraviolet energy range and much larger than its band gap value, which can be due to dipole forbidden inter-band transitions.

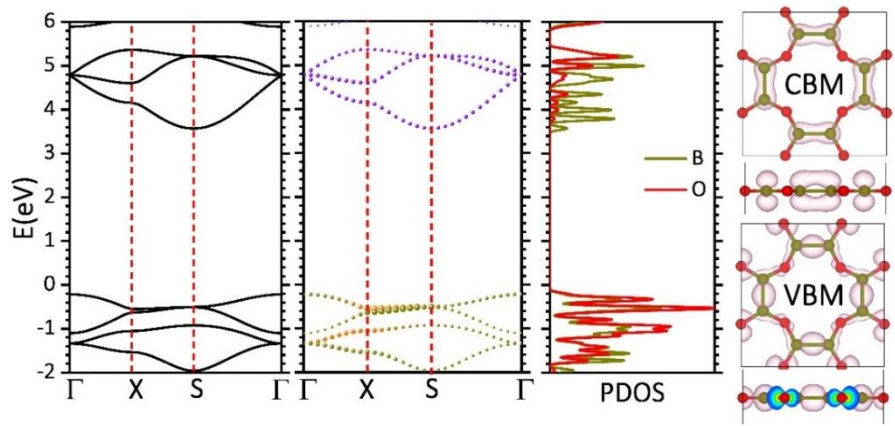

Fig. 2. HSE06 band structure, orbital projected band structure, atom-type partial density of states (PDOS), and charge density distribution of the valence band maximum (VBM) and conduction band minimum (CBM) of the BO monolayer. The Fermi level is set to 0 eV.

It is well-known that the electronic structure of layered materials might be effectively modulated by thickness due to well-known quantum conferment effects. To assess the electronic behavior of the BO nanosheets with different number of layers, we first constructed four different staggered and eclipsed configurations for the BO bilayer (2L-BO). It is found that the most staggered stacking configuration in which the six membered rings ($B_4O_2$) of each layer are in front of the pores of the other layer exhibits the lowest energy. This configuration is by 0.45 eV/unitcell more stable than the eclipsed configuration. The most staggered configuration was also used to construct the BO trilayer (3L-BO). The HSE06 band structures of the 2L- and 3L-BO are shown in Fig. S2. It can be clearly seen that the general features of these band structures are close to those of the BO monolayer, with a minor difference that the CBM in 2L- and 3L-BO is slightly off S toward X point. The calculated band gaps are 3.80 and 3.78 eV for 2L- and 3L-BO, respectively, which are only slightly different than that of the BO monolayer, indicating very weak out of plane quantum confinement in the BO nanosheets. This observation can be simply explained by considering the point that VB states of each of stacked monolayers consists in-plane σ character, leading to an ineffective interlayer (bonding-anti-bonding) interactions. The CB states of each monolayer although preserving the π-character, their interlayer interaction is limited because of the staggered stacking. This observation results in the situation that VBs and CBs of 2L- and 3L-BO occur at the same energy ranges as those in the pristine BO monolayer. The aforementioned finding can be supported by: (1) finding similar absolute band edge positions for the BO monolayer to trilayer systems (VBM(CBM) of the monolayer: -6.88 (-3.1) eV; 2L: -6.89 (-3.09) eV and 3L: -6.89(-3.11) eV), (2) finding a much smaller band gap of 3.19 eV for the BO bilayer in the eclipsed stacking mode (Fig. S2), compared to that of the BO monolayer.

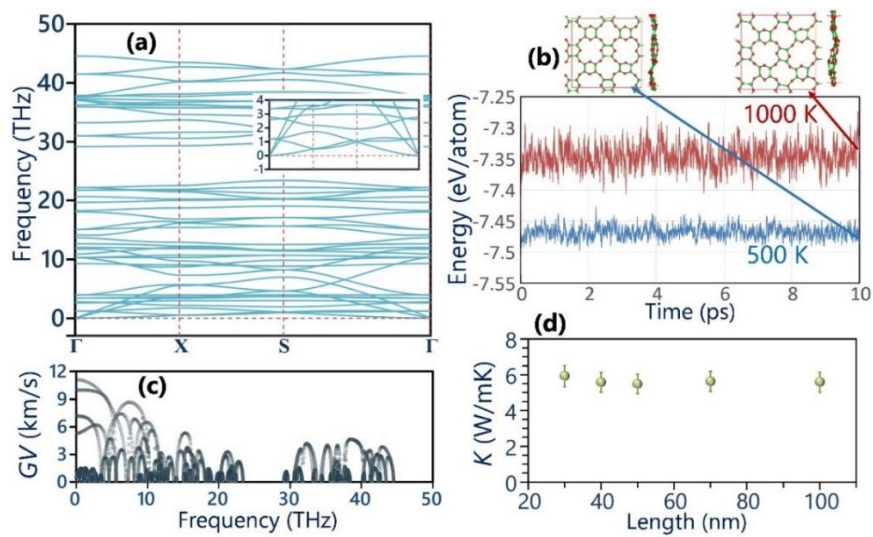

**Fig. 3**. (a) Phonon dispersion relation of the BO monolayer, here inset shows the zoomed view of low-frequency phonon modes. (b) The AIMD results for the energy per atom of the BO nanosheet at 500 and 1000 K alongside with the top and side views of the atomic lattices after the 10 ps long calculations. (c) Predicted phonons' group velocity (*GV*) of the single-layer BO. (d) Length dependent phononic thermal conductivity of the BO monolayer at the room temperature.

Having elaborately studied the structural and electronic characteristics, we now shift our consideration to the analysis of the dynamical and thermal stability of the BO monolayer on the basis of the phonon spectra and AIMD results, respectively. In Fig. 3a, the predicted phonon dispersion curve of the BO monolayer with the aid of the fitted MLIP is presented, which reveals that none of the three acoustic phonon and entire optical branches exhibit imaginary frequencies, indicating the desired dynamical stability of the suspended BO system. AIMD simulations results carried out at 500 and 1000 K for 10 ps shown in Fig. 3b, clearly confirm that the BO monolayer exhibits complete intactness and remarkable flexibility at the considered temperatures. The AIMD and phonon spectra results demonstrate the robust thermal and dynamical stability of the suspended BO monolayer [31–34]. As shown in Fig. 3a inset, from the Γ point three acoustic phonon modes originate, in which the out-of-plane (ZA) shows quadratic relations, whereas the two in-plane counterparts appear with almost linear dispersions, in analogy with those in the single-layer graphene and most of 2D materials [18]. The in-plane longitudinal acoustic phonon modes show the sharpest dispersions, which reveal their largest phonon group velocities around 10.4 km/s, as shown in Fig. 3c. The transverse acoustic phonon modes show considerably lower maximum group velocities of around 7.1 km/s. It is noticeable that the ZA acoustic and the majority of optical phonon modes exhibit narrow dispersions, which reveal their remarkably suppressed phonon group velocities, consistent with results shown in Fig. 3c. In addition, except the ZA mode, significant band crossing is exposed for the all phonons throughout the whole frequency range, revealing remarkable scattering rates and short lifetimes for the majority of phononic energy carriers in the BO monolayer. From the presented phonon dispersion relation, the low group velocities and short lifetimes can be concluded for the phonons in the BO monolayer, both suggesting low lattice thermal conductivity in this novel nanosheet. In Fig. 3d, the MLIP-based NEMD predictions for the length effects on the room temperature phononic thermal conductivity of the BO monolayer is plotted, assuming a thickness of the 4.19 Å according to the experimental report [5]. As it is clear and consistent with our preliminary findings, the thermal conductivity of the system with the length of 30 nm is completely converged to a value around 5.6 ±0.5 W/mK, as compared with other samples with longer lengths up to 100 nm. It appears that the thermal conductivity of the BO nanosheets is by almost three, two and one orders of magnitude lower than that of the graphene [35], h-BN [36] and phosphorene [37], respectively, which confirm that it belongs to the family of low thermal conductivity 2D material. The predicted low thermal conductivity of the BO nanosheet is nevertheless rather high to be promising as a thermoelectric material, and also reveals that it may face thermal management issues when employed in nanoelectronics.

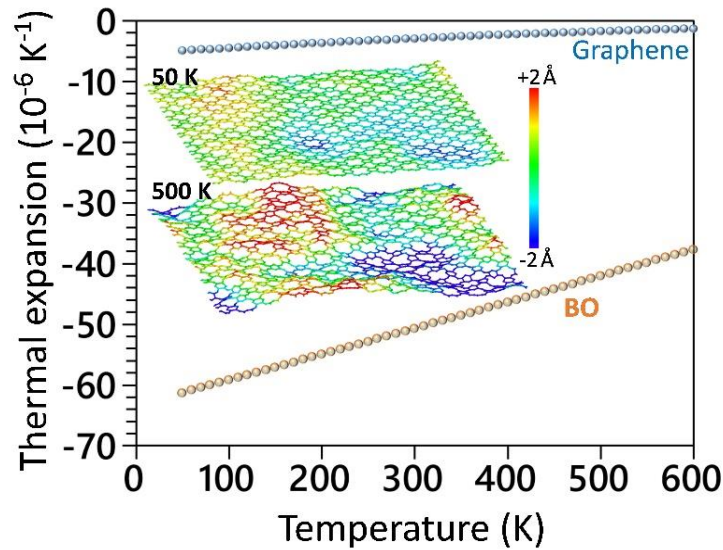

Fig. 4, MLIP-based molecular dynamics results for the thermal expansion coefficients of the suspended BO and graphene [38] monolayers as a function of temperature. Insets illustrate the BO monolayers equilibrated at 50 and 500 K, with color coding represents the out-of-plane displacements of atoms.

It is also useful to examine the thermal expansion behaviour of the BO monolayer using the developed MLIP-based model, using a validated as discussed in our recent study [38]. The temperature dependency of the thermal expansion coefficients was evaluated based on the $\frac{1}{A}\frac{dA}{dT}$ relation, where $A$ denotes the projected area at a given temperature of $T$ [38,39]. To obtain the aforementioned relation, eight independent calculations with uncorrelated initial velocities were carried out over a 12×12×1 supercell, the projected areas at every temperatures were averaged, and a polynomial curve was fitted to report the thermal expansion coefficients [38]. The details of the MTP-based molecular dynamics calculations are fully given in our earlier study [38]. In Fig. 4, the thermal expansion coefficients of the single-layer BO are compared with those of the graphene [38]. The thermal expansion coefficients of the suspended BO, graphene and h-BN monolayers at room temperature were predicted to be -50.7, −2.9 [38] and -7.8 ×10$^{-6}$ K$^{-1}$ [38], respectively. Our results confirm an almost 17 folds larger negative thermal expansion of the BO nanosheet than the graphene counterpart under ambient condition. As illustrated in the Fig. 4 insets for the atomic structures of the BO lattice at 50 and 500 K, it is noticeable that at low temperatures this nanosheet presents a fairly flat configuration, which however at higher temperatures due to presence of nanoporous lattice and low bending rigidity undergoes substantial out-of-plane wrinkles. The substantial formation of wrinkles at higher temperatures results in the shrinkage of the projected area of the sheet and subsequently appears with ultrahigh negative thermal expansion coefficients [38]. As a very interesting finding, despite of much lower porosity of the BO nanosheet as compared with graphyne counterparts, the BO lattice can show larger negative thermal expansion coefficients at room temperature than the majority of graphyne monolayers [40].

We finally explore the mechanical properties of the BO monolayer, with the DFT and also the fitted MLIP-based molecular dynamics model. According to the experimental measurements [5], a fixed thickness of 4.19 Å was assumed during the entire mechanical loading to report the stress values in the GPa unit. As for the validation, the $C_{11}$($C_{12}$) values of the single-layer BO were predicted to be 142(56) GPa and 138(59) GPa, by the DFT and MLIP-based model, respectively, which confirms the excellent accuracy of the developed classical molecular model. In Fig. 5a, the uniaxial stress-strain curves of the BO monolayer by the DFT at the ground state and MLIP-based model carried out at 1 K are plotted, which furthermore confirm the excellent accuracy of the developed MLIP in the analysis of the mechanical properties of the BO nanosheet. It is also noticeable from the MLIP-based results that after reaching the ultimate tensile strength, the stress values drop sharply, which is an indication of the brittle failure mechanism, consistent with well-known behavior of this system [5]. Nonetheless, as previously discussed for the case of graphene [14,16], with the ground state DFT calculations because of not taking into account the atomic vibrations, the drop of the stress values of the BO monolayer appears with a smooth pattern and thus making it controversial to conclude about the failure mechanism. From Fig. 5a inset it is also clear that the failure in this system occurs along the homonuclear B-B bonds. The uniaxial stress-strain responses of the BO monolayer at 300 and 500 K by the validated MLIP-based model are also presented in Fig. 5b, which as expected reveal suppression of the ultimate tensile strength by increasing the ambient temperature. The elastic modulus and tensile strength of the BO monolayer at room temperature are predicted to be 107 and 25 GPa, respectively. It is worth mentioning that according to the Griffth theory [41], the elastic modulus to ultimate tensile strength ratio of a defect-free material should be around 10. This ratio for the pristine graphene, $MoS_2$, and silicene monolayers were reported to be 8 [42], ~11 [43], and ~9 [44] respectively. As an unexpected behaviour, the aforementioned ratio for the BO monolayer is noticeably predicted to be around 4, which is almost half of that for graphene. From the 1 to 500 K, the tensile strength of the BO monolayers drops by around 16%, which also confirms the unique ability of this novel nanosheet to exhibit decent tensile stability at elevated temperatures.

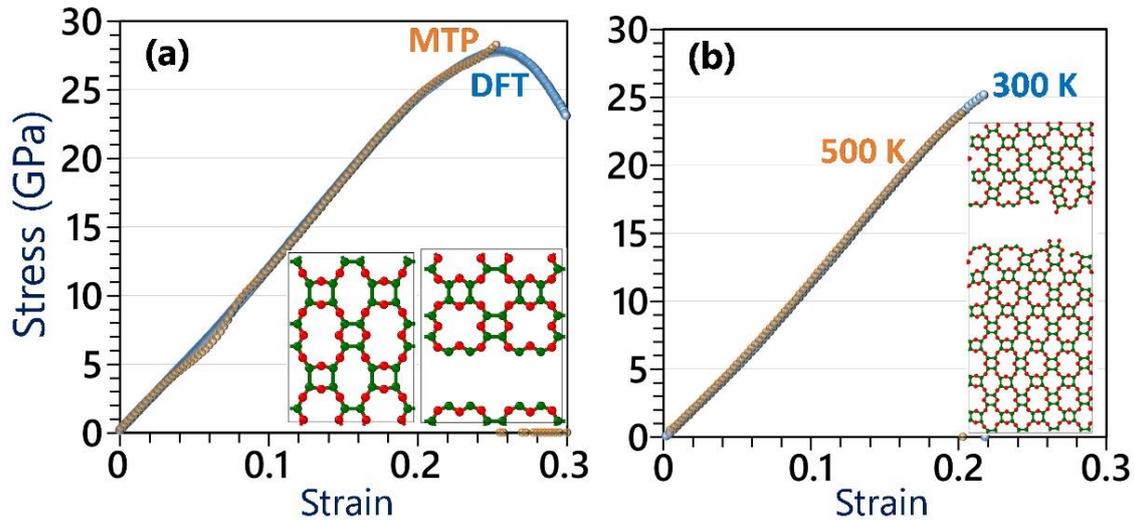

Fig. 5, (a) Comparison of the uniaxial stress-strain curves of the BO monolayer by the MTP simulations conducted at 1 K with the ground state DFT. (b) Predicted uniaxial stress-strain curves of the BO monolayer at 300 and 500 K. Insets show the atomic configurations during the uniaxial loading illustrated using the OVITO package [45].

4. Concluding remarks

Inspired by the recent experimental discovery on the layered atomic structure of the BO material synthesized by the thermal treatment of tetrahydroxydiboron by Perras *et al.* [5], herein the stability and physical properties of the suspended BO monolayers were explored using the state-of-the-art theoretical calculations. DFT results not only confirm the decent thermal stability of the BO nanosheets but also reveal that the formation of BO nanomembranes through the condensation reaction of tetrahydroxydiboron molecules is thermodynamically unfavorable by +2.35 eV/unitcell. The HSE06 calculations predict that the BO monolayer is an indirect semiconductor with a large band gap of 3.78 eV. The optical absorption of the pristine single-layer BO is nonetheless found to be shifted toward the near UV region due to the dipole forbidden inter-band transitions. The dynamical stability, thermal expansion and conductivity, and mechanical properties of the BO monolayer were explored using the MLIPs. The room temperature phononic thermal conductivity of the BO monolayer is found to be relatively low, 5.6 ±0.5 W/mK, due to suppressed phonons' group velocities and lifetimes. The thermal expansion coefficient of the suspended BO monolayer at room temperature was predicted to be by almost 17 folds larger than that of the graphene. The elastic modulus (tensile strength) of the single-layer BO were predicted to be 120 (27.8) GPa and 114 (28.3) GPa, by the DFT and MLIP-based model, respectively, which confirms the excellent accuracy of the developed classical model. The failure mechanism of the BO nanosheets is found to be brittle, occurring with the debonding of homonuclear B-B bonds. The elastic modulus to ultimate tensile strength ratio of the BO monolayer at room temperature is predicted to be around 4, which is almost half of that for graphene and other prominent 2D materials. This study highlights the anomalous tensile strength to elastic modulus ratio, ultrahigh negative thermal expansion coefficients, low lattice thermal conductivity and indirect wide band gap characteristics of the boron monoxide monolayer.

## Declaration of competing interest

The authors declare that they have no known competing financial interests or personal relationships that could have appeared to influence the work reported in this paper.

## Data availability

The energy-minimized lattice in the VASP and LAMMPS input formats, and the fitted MTP are available to download as a zip file. Additional data presented in this study are also available on request from the corresponding author.

## Acknowledgment

B. M. and X. Z. appreciate the funding by the Deutsche Forschungsgemeinschaft (DFG, German Research Foundation) under Germany's Excellence Strategy within the Cluster of Excellence PhoenixD (EXC 2122, Project ID 390833453). F.S. thanks the Persian Gulf University Research Council, Iran, for the support of this study. B. M is greatly thankful to the VEGAS cluster at Bauhaus University of Weimar the cluster system team at the Leibniz University of Hannover, for providing the computational resources.

Supporting information

# Anomalous tensile strength and thermal expansion, and low thermal conductivity in wide band gap boron monoxide monolayer

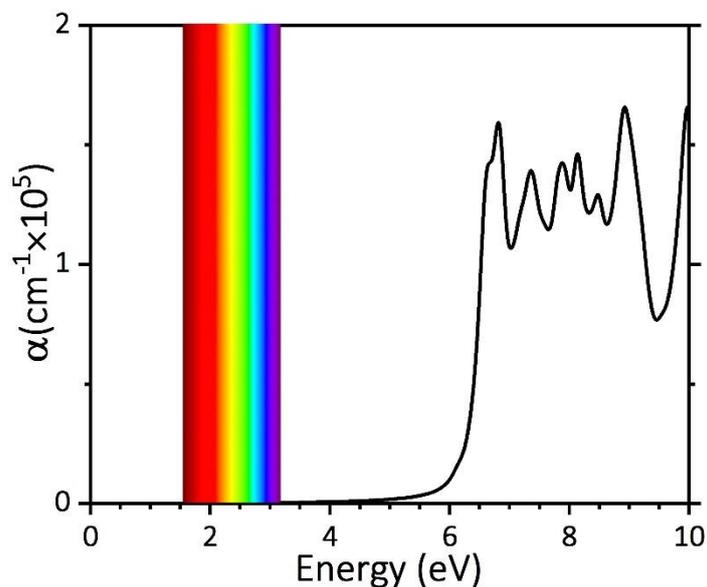

**Fig. S1.** Frequency dependent absorption coefficients for BO monolayer, calculated using the HSE06 functional. The plotted result is for the light incident from z direction and polarized along [100] and [010] directions. The visible light energy range is shown using the rainbow color bar.

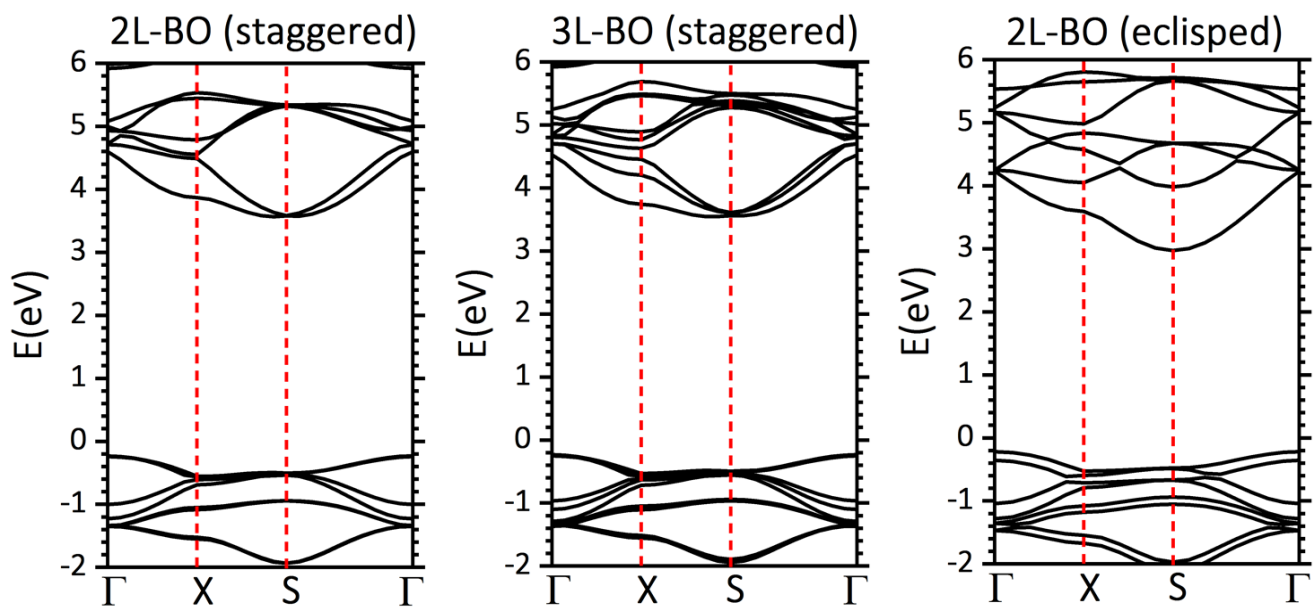

**Fig. S2.** HSE06 band structure of BO bilayer and trilayer in staggered stacking mode and BO bilayer in eclipsed stacking mode. The Fermi level is set to 0 eV.